# Fully Few-shot Class-incremental Audio Classification Using Multi-level Embedding Extractor and Ridge Regression Classifier

*Yongjie Si[1], Yanxiong Li[1], Jiaxin Tan[1], Qianhua He[1], Il-Youp Kwak[2]*

[1]School of Electronic and Information Engineering, South China University of Technology, Guangzhou, China
[2] Department of Statistics and Data Science, Chung-Ang University, Republic of Korea
eeyongjiesi@mail.scut.edu.cn, eeyxli@scut.edu.cn, tanjiaxin02@126.com, eeqhhe@scut.edu.cn, ikwak2@cau.ac.kr

## Abstract

In the task of Few-shot Class-incremental Audio Classification (FCAC), training samples of each base class are required to be abundant to train model. However, it is not easy to collect abundant training samples for many base classes due to data scarcity and high collection cost. We discuss a more realistic issue, Fully FCAC (FFCAC), in which training samples of both base and incremental classes are only a few. Furthermore, we propose a FFCAC method using a model which is decoupled into a multi-level embedding extractor and a ridge regression classifier. The embedding extractor consists of an encoder of audio spectrogram Transformer and a fusion module, and is trained in the base session but frozen in all incremental sessions. The classifier is updated continually in each incremental session. Results on three public datasets show that our method exceeds current methods in accuracy, and has advantage over most of them in complexity. The code is at https://github.com/YongjieSi/MAR.

## 1. Introduction

Audio classification (AC) is a task to classify different types of sounds. It is a key module for implementing many audio or video processing tasks, such as acoustic event detection [1]-[5], acoustic scene classification [6]-[9], video analysis [10]-[12], speaker analysis [13]-[15], and keyword spotting [16]-[19].

Many efforts are made on AC. According to the abilities to identify old and new classes and the required numbers of training samples, the existing methods can be divided into five types, namely Many-shot AC (MAC), Few-shot AC (FAC), Class-incremental AC (CAC), FCAC, and FFCAC. The MAC methods [20]-[24] can identify predefined classes with abundant training samples per class but lack the ability to classify new classes. The FAC methods [25]-[28] can identify new classes using few training samples per class, but cannot memorize old classes. The CAC methods [29]-[31] is able to memorize old classes and continually identify new classes. Their deficiency is that sufficient training samples per class are required for model training in all sessions. The FCAC methods [32]-[37] can continually identify new classes with few training samples and meanwhile memorize old classes. However, they require a substantial number of training samples per base class to train a strong model in base session. To reduce requirement for the number of training samples per base class, FFCAC methods are proposed [38], [39] which can identify incremental classes with memorizing base classes using only few training samples of all classes. However, their complexity is very high and their accuracy needs to be further improved.

To overcome the above shortcomings of AC methods, we propose a FFCAC method using a Multi-level Embedding Extractor (MEE) and a Ridge Regression Classifier (RRC) in this paper. To compare the performance of different methods, experiments are conducted on three public datasets (FSC-89, NSynth-100 and LS-100 [39]). Main contributions of this work are briefly summarized as follows.

1. We design a MEE which consists of an encoder of Audio Spectrogram Transformer (AST) [40] and a fusion module. The proposed MEE can extract discriminative embeddings by fusing features output by multiple blocks of the AST encoder.

2. We design a RRC which can reduce the correlations between various dimensions of embeddings. As a result, the overfitting issue caused by few training samples is expected to be alleviated. The RRC weights are updated by an analytical solution without being retrained in each incremental session.

3. We propose a FFCAC method using a model which is decoupled into a MEE and a RRC. Results show that our method exceeds previous methods in Average Accuracy (AA) and Performance Degradation rate (PD), and has advantage over most of them in complexity.

## 2. Method

In this section, we describe our method in detail, including problem definition, method framework, multi-level embedding extractor, and ridge regression classifier.

### 2.1. Problem Definition

The FFCAC problem has two types of sessions: base session (Session 0) and incremental session (Sessions 1 to $M$). $M$ is total number of incremental sessions. The training and testing datasets of different sessions are denoted by $\{\boldsymbol{D}_0^{tr}, \boldsymbol{D}_1^{tr}, ..., \boldsymbol{D}_m^{tr}, ..., \boldsymbol{D}_M^{tr}\}$ and $\{\boldsymbol{D}_0^{te}, \boldsymbol{D}_1^{te}, ..., \boldsymbol{D}_m^{te}, ..., \boldsymbol{D}_M^{te}\}$, respectively. $\boldsymbol{D}_m^{tr}$ and $\boldsymbol{D}_m^{te}$ have the same label set which is represented by $Y_m$. The datasets of different sessions have different types of classes, namely $\forall m', m,$ and $m' \neq m$, $Y_{m'} \cap Y_m = \phi$. In the $m$th session, only $\boldsymbol{D}_m^{tr}$ can be used to update the model, and the updated model needs to be evaluated on the testing datasets of current and all prior sessions, namely $\boldsymbol{D}_0^{te} \cup \boldsymbol{D}_1^{te} ... \cup \boldsymbol{D}_m^{te}$. Each training dataset $\boldsymbol{D}_m^{tr}$ ($0 \leq m \leq M$) is a small-scale dataset which has $N_m$ classes and $K_m$ samples per class.

However, in the FCAC problem, $\boldsymbol{D}_0^{tr}$ is a large-scale dataset with abundant samples per class and $\boldsymbol{D}_m^{tr}$ ($1 \leq m \leq M$) is a small-scale dataset with few samples per class. That is, the only difference between the FCAC and FFCAC problems lies in the size of $\boldsymbol{D}_0^{tr}$.

### 2.2. Method Framework

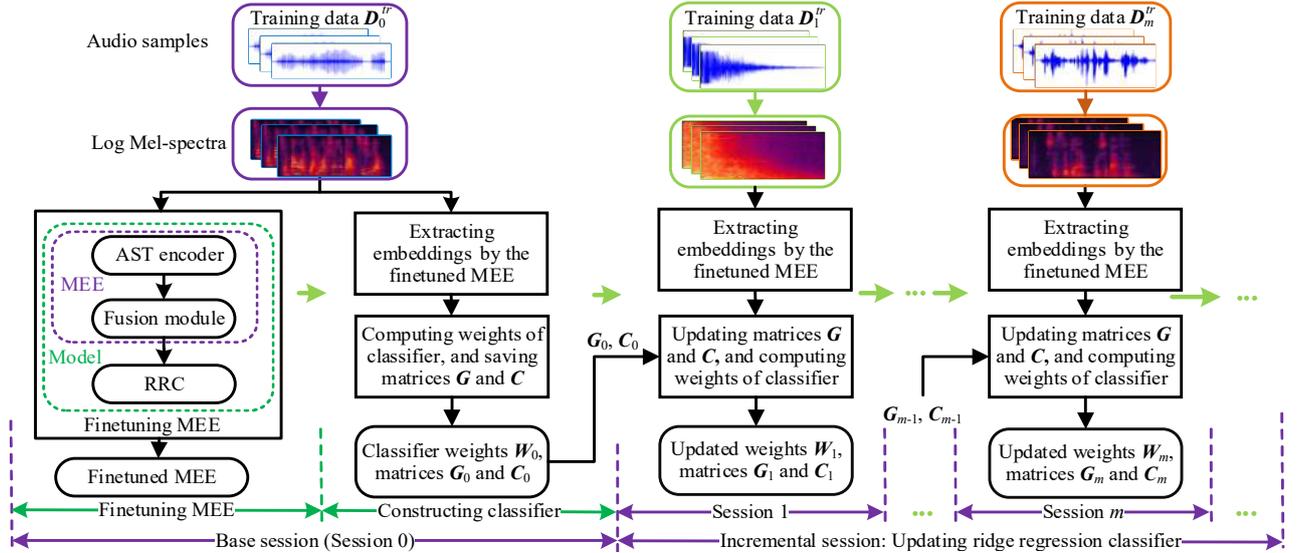

Figure 1. *Framework of the proposed method. AST: Audio Spectrogram Transformer; MEE: Multi-level Embedding Extractor; RRC: Ridge Regression Classifier; **G** and **C**: the matrices defined by Eq. 6 and 7, respectively; $W_m$: classifier weights in session m.*

As shown in Figure 1, the framework of our method consists of a base session and $M$ incremental sessions. The model consists of a MEE and a RRC. The MEE comprises an AST encoder and a fusion module. The AST encoder is initialized with the parameters of a pretrained AST. The classifier weights are computed by a closed-form solution of regularized least squares. The base session includes two steps: finetuning MEE and constructing classifier. In each incremental session, the RRC is continually updated for recognizing all seen classes.

In the base session, Log Mel-Spectrum (LMS) [41] is first extracted from each sample to finetune the MEE under the supervision of loss

$$\ell(e_k, y_k) = -log \frac{\exp(\eta \cos(e_k, W_{y_k}))}{\exp(\eta \cos(e_k, W_{y_k})) + \sum_{y_j \neq y_k} \exp(\eta \cos(e_k, W_{y_j}))}, \quad (1)$$

where $\eta$, $W_{y_k}$ and $e_k$ denote a scale factor, the classifier weights for class $y_k$ and the embedding of the $k$th sample, respectively; $\cos(\cdot, \cdot)$ is a cosine function. The embeddings are extracted from the LMS of training samples with the finetuned MEE and used to compute the classifier weights. The matrices **G** (defined by Eq. 6) and **C** (defined by Eq. 7) are saved for updating the classifier in incremental sessions.

In the $m$th ($1 \leq m \leq M$) incremental session, the MEE is frozen and used to extract the embeddings from the LMS of samples. The embeddings are used to update matrices **G** and **C**, and to compute the classifier weights. During inference, the cosine similarity between the embedding of each testing sample and classifier weights is used to classify the testing sample.

### 2.3. Multi-level Embedding Extractor

Figure 2 depicts the structure of the MEE which includes an operation of patch split, an AST encoder and a fusion module. Each LMS is divided into $Z$ sub-spectra along the time and frequency axes with a step size of $d$. The size of a LMS and a sub-spectrum are $S_f \times S_t$ and $s_f \times s_t$, respectively. $S_f$ (or $s_f$) and $S_t$ (or $s_t$) denote the LMS's (or sub-spectrum's) dimensions of frequency and time, respectively. $Z$ is computed by

$$Z = \left\lfloor \frac{S_f - s_f}{d} + 1 \right\rfloor \times \left\lfloor \frac{S_t - s_t}{d} + 1 \right\rfloor, \quad (2)$$

where $\lfloor \cdot \rfloor$ is a floor function. $Z$ sub-spectra are sequentially fed to AST encoder which is pre-trained using abundant samples. Each block of the AST encoder has the same structure, and consists of an attention sub-block and a feed-forward sub-block.

The attention sub-block is similar to the self-attention module of the Transformer [42], and includes the operations of linear transformations, matrix multiplication, scale and Softmax in sequence. The feed-forward sub-block is composed of the operations of element-wise addition, layer normalization and Fully-Connected (FC) layer in sequence.

The motivation for designing the structure of MEE is that features extracted by different blocks of the AST encoder focus on different levels of time-frequency characteristics of samples. The shallower blocks extract concrete and local characteristics, while the deeper blocks extract abstract and global characterisitcs. The MEE is designed with a fusion module to incoporate multi-level features into the embedding. Hence, the extracted embedding is expected to represent each sample well and to be effectively distinguish various classes.

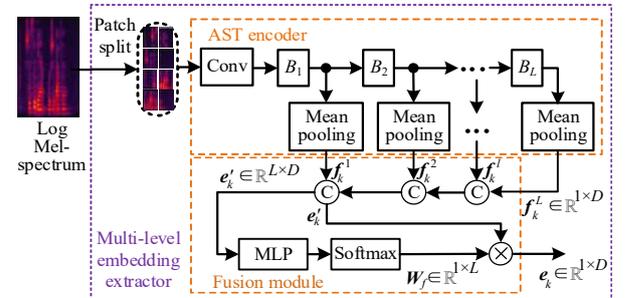

Figure 2: *MEE structure. Conv: convolution layer; MLP: Multi-Layer Perceptron; $B_1$ to $B_L$: blocks of AST encoder; $e'_k$ ($e_k$): embeddings before (after) fusion for the kth sample. $f_k^l$: feature output by the lth ($1 \leq l \leq L$) block for the kth sample; $W_f$: weight matrix; L: no. of blocks; D: dimension of embeddings; ©: concatenation; ⊗: multiplication.*

As shown in Figure 2, the features output by each block are concatenated and fed to the fusion module which consists of a Multi-Layer Perceptron (MLP), Softmax layer, and operation of matrix multiplication. The embeddings are computed by

$$e_k = W_f \otimes e'_k \quad (3)$$
$$W_f = Softmax(h_{\theta_2}(ReLU(h_{\theta_1}(e'_k)))) \quad (4)$$

where $e'_k$ and $e_k$ denote embeddings before and after fusion for the $k$th sample, respectively. $W_f$ denotes the weight matrix; $h_{\theta_1}(\cdot)$ and $h_{\theta_2}(\cdot)$ are transformation of the input and output layers of the MLP with parameters of $\theta_1$ and $\theta_2$, respectively.

### 2.4. Ridge Regression Classifier

In incremental learning, there are two types of classifiers: parametric classifier (e.g., linear classifier) and non-parametric classifier (e.g., prototype classifier) [43]. The weights of the former need to be updated by iterative gradient, while weights of the latter can be directly computed by closed-form formula with lower computational complexity. The non-parametric classifier has been widely adopted in tasks of class-incremental learning [44]. Since the diversity and under-representation of samples, the relationships between various dimensions of the feature (e.g., embedding) of each sample are overemphasized. Hence, correlations between various dimensions of the feature are significant and the trained model is prone to overfitting [45]. Ridge regression can reduce the correlations between various dimensions of feature via minimizing the least mean square error with $l_2$-regularization [46]. We use the RRC to reduce the correlations between various dimensions of the feature for mitigating the overfitting problem in the FFCAC. Specifically, the RRC weights are computed by

$$W = (E^T E + \lambda I)^{-1} E^T Y \quad (5)$$
$$G = E^T E \quad (6)$$
$$C = E^T Y \quad (7)$$

where $W$, $E$, $Y$, $I$ and T are the classifier weights, concatenated embeddings, one-hot encoded labels, identity matrix and transpose operation, respectively. $G$ and $C$ are two matrices for updating classifier weights. $\lambda$ is the regression parameter which is obtained by cross-validation using training samples. In the $m$th ($1 \leq m \leq M$) session, $G$ and $C$ are updated by

$$G_m = G_{m-1} + E_m^T E_m \quad (8)$$
$$C_m = C_{m-1} + E_m^T Y_m \quad (9)$$

where $E_m$ and $Y_m$ denote the concatenated embeddings and one-hot encoded labels of training samples in session $m$, respectively. $\{G_m, C_m\}$ and $\{G_{m-1}, C_{m-1}\}$ are the matrices in session $m$ and $m$-1, respectively.

## 3. Experiments

### 3.1. Experimental Datasets

Table 1 lists detailed information of the LS-100, FSC-89, and NSynth-100. The three datasets have been widely adopted for the task of AC and can be obtained from the three websites:
https://www.modelscope.cn/datasets/pp199124903/LS-100/summary;
https://www.modelscope.cn/datasets/pp199124903/FSC-89/summary;
https://www.modelscope.cn/datasets/pp199124903/NSynth-100/summary.

We randomly select samples of fifty classes from each one of the three datasets. The selected samples are split into ($M$+1) parts without overlaps of classes, namely $D_m$ ($0 \leq m \leq M$). $D_m$ consists of training dataset $D_m^{tr}$ and testing dataset $D_m^{te}$.

Table 1: *Detailed information of LS-100/ NSYNTH-100/FSC-89*

| Parameters | $D_m$ | |
|---|---|---|
| | $D_m^{tr}$ | $D_m^{te}$ |
| #Classes | 100/100/89 | 100/100/89 |
| #Samples | 50000/20000/44500 | 10000/10000/17800 |
| Length (hours) | 27.77/22.24/12.36 | 5.55/10.00/4.95 |
| #Samples/Class | 500/200/500 | 100/100/200 |

#Samples/Class: number of samples per class; $D_m$: samples of classes in all sessions.

### 3.2. Experimental Setup

AA and PD have been widely adopted as the performance metrics in previous works. They are defined by

$$AA = \frac{1}{M+1} \sum_{m=0}^{M} A_m, \quad (10)$$
$$PD = A_0 - A_M, \quad (11)$$

where $A_m$ stands for the accuracy in session $m$.

In each training step, $N \cdot K$ samples ($N$ classes and $K$ samples per class) from the $D_m^{tr}$ are used to train the model. All testing samples from all classes so far are fed to the model for testing. The final result is the average of 100 tests. Table 2 presents the settings of major parameters of our method.

Table 2: *Settings of major parameters of our method*

| Parameters | Values | Parameters | Values |
|---|---|---|---|
| Frame length/shift | 25/15 ms | Learning rate | 0.001 |
| Dimension of LMS, $S_f$ | 128 | Weight decay | 0.0005 |
| Dimension of embeddings, $D$ | 768 | Training epochs | 100 |
| $N, K$ | 5, 5 | Scale factor $\eta$ | 16 |

### 3.3. Ablation Study

Ablation experiments are done to validate the contributions of the fusion module of the MEE and the classifier (Prototype Based Classifier (PBC) [33] or RRC). Table 3 lists the results obtained by our method with various combinations of the fusion module and the classifier on the NSynth-100. In cases ① and ②, the PBC is used as the classifier. In cases ① and ③, the fusion module is not used and the feature $f_k^L$ output by the last block of the AST encoder is fed to the classifier. Based on the results in all the four cases, it can be known that the fusion module and the classifier have contributions to the performance improvement. When the fusion module and the RRC are used together (case ④), our method obtains the best result with the highest AA scores and the lowest PD scores.

Table 3: *Ablation Results obtained by our method with different combinations of the fusion module and classifier on the NSynth-100*

| Cases | Fusion module | Classifier | Accuracy scores in various sessions (%) | | | | | | | | | | AA (%) | PD (%) |
|---|---|---|---|---|---|---|---|---|---|---|---|---|---|---|
| | | | 0 | 1 | 2 | 3 | 4 | 5 | 6 | 7 | 8 | 9 | | |
| ① | × | PBC | 71.80 | 66.10 | 59.40 | 55.50 | 51.72 | 45.93 | 41.94 | 42.88 | 39.67 | 40.02 | 51.50 | 31.78 |
| ② | √ | PBC | 76.00 | 72.80 | 67.53 | 65.20 | 61.28 | 54.30 | 50.97 | 52.70 | 48.02 | 49.14 | 59.79 | 26.86 |
| ③ | × | RRC | 77.00 | 73.80 | 72.13 | 70.16 | 69.48 | 62.01 | 59.86 | 60.43 | 54.84 | 53.40 | 65.31 | 23.60 |
| ④ | √ | RRC | **80.78** | **83.02** | **74.23** | **70.85** | **70.25** | **62.77** | **59.90** | **66.79** | **59.86** | **58.07** | **68.65** | **22.71** |

Table 4: *Results obtained by different methods on the LS-100*

| Methods | Accuracy scores in various sessions (%) | | | | | | | | | | AA (%) | PD (%) |
|---|---|---|---|---|---|---|---|---|---|---|---|---|
| | 0 | 1 | 2 | 3 | 4 | 5 | 6 | 7 | 8 | 9 | | |
| Finetune [47] | 91.25 | 49.29 | 35.42 | 24.26 | 23.37 | 18.95 | 20.08 | 16.88 | 15.75 | 13.34 | 30.86 | 77.91 |
| iCaRL [48] | 91.32 | 52.74 | 37.22 | 26.19 | 25.49 | 24.03 | 25.67 | 25.06 | 23.96 | 23.03 | 35.47 | 68.29 |
| DFSL [32] | 90.56 | 59.02 | 46.18 | 39.55 | 37.41 | 34.46 | 35.79 | 34.58 | 32.96 | 32.90 | 44.34 | 57.66 |
| ARP [35] | 90.26 | 65.81 | 50.39 | 44.07 | 43.04 | 39.91 | 40.25 | 39.16 | 37.52 | 37.03 | 48.74 | 53.23 |
| CEC [49] | 91.33 | 68.52 | 53.32 | 45.47 | 44.39 | 41.65 | 42.42 | 41.35 | 39.61 | 39.63 | 50.77 | 51.70 |
| FACT [50] | 90.79 | 68.15 | 53.00 | 45.35 | 44.24 | 41.50 | 42.18 | 41.11 | 39.33 | 39.29 | 50.49 | 51.50 |
| PAN [33] | 91.44 | 67.18 | 51.60 | 45.06 | 44.07 | 41.16 | 41.63 | 40.65 | 39.00 | 38.78 | 50.06 | 52.66 |
| AMFO [37] | 92.88 | 69.98 | 53.72 | 46.88 | 45.92 | 42.78 | 43.12 | 42.51 | 40.77 | 40.35 | 51.89 | 52.53 |
| AISP [39] | 94.48 | 71.58 | 56.28 | 47.81 | 46.29 | 45.27 | 47.62 | **47.90** | **47.13** | 48.60 | 55.30 | 45.88 |
| Ours | **94.50** | **76.85** | **62.10** | **55.51** | **52.98** | **48.56** | **48.33** | 46.37 | 46.93 | **57.36** | **58.95** | **37.14** |

Table 5: *Results obtained by different methods on the NSynth-100*

| Methods | Accuracy scores in various sessions (%) | | | | | | | | | | AA (%) | PD (%) |
|---|---|---|---|---|---|---|---|---|---|---|---|---|
| | 0 | 1 | 2 | 3 | 4 | 5 | 6 | 7 | 8 | 9 | | |
| Finetune [47] | 73.21 | 57.24 | 46.97 | 44.90 | 43.73 | 34.23 | 29.46 | 21.68 | 21.89 | 22.68 | 39.60 | 50.53 |
| iCaRL [48] | 72.99 | 59.70 | 55.29 | 49.88 | 48.31 | 41.33 | 40.56 | 39.99 | 39.52 | 40.15 | 48.77 | 32.84 |
| DFSL [32] | 72.10 | 65.98 | 58.12 | 53.77 | 51.55 | 45.58 | 42.18 | 42.87 | 39.28 | 40.44 | 51.19 | 31.66 |
| ARP [35] | 72.76 | 61.37 | 57.59 | 53.81 | 51.53 | 46.10 | 42.75 | 43.91 | 40.13 | 40.92 | 51.09 | 31.84 |
| CEC [49] | 72.67 | 68.20 | 62.27 | 58.11 | 56.20 | 50.20 | 46.51 | 47.81 | 43.77 | 44.00 | 55.08 | 29.67 |
| FACT [50] | 72.95 | 66.85 | 61.40 | 57.36 | 54.99 | 49.10 | 45.55 | 46.76 | 42.90 | 43.97 | 54.18 | 28.98 |
| PAN [33] | 72.16 | 66.87 | 61.14 | 56.93 | 55.02 | 49.08 | 45.47 | 46.90 | 42.96 | 43.08 | 53.96 | 29.08 |
| AMFO [37] | 74.98 | 70.6 | 64.5 | 60.93 | 57.32 | 51.47 | 46.54 | 48.67 | 45.01 | 45.57 | 56.56 | 29.41 |
| AISP [39] | 77.20 | 72.85 | 66.58 | 63.74 | 62.27 | 56.42 | 53.90 | 54.84 | 53.22 | 52.31 | 61.33 | 24.89 |
| Ours | **80.78** | **83.02** | **74.23** | **70.85** | **70.25** | **62.77** | **59.9** | **66.79** | **59.86** | **58.07** | **68.65** | **22.71** |

Table 6: *Results obtained by different methods on the FSC-89*

| Methods | Accuracy scores in various sessions (%) | | | | | | | | | | AA (%) | PD (%) |
|---|---|---|---|---|---|---|---|---|---|---|---|---|
| | 0 | 1 | 2 | 3 | 4 | 5 | 6 | 7 | 8 | 9 | | |
| Finetune [47] | 53.45 | 33.76 | 28.47 | 24.28 | 18.63 | 15.05 | 15.32 | 14.31 | 13.62 | 12.94 | 22.98 | 40.51 |
| iCaRL [48] | 53.47 | 30.61 | 28.75 | 25.24 | 19.29 | 16.90 | 18.21 | 17.89 | 18.55 | 17.53 | 24.64 | 35.94 |
| DFSL [32] | 52.65 | 36.09 | 34.64 | 31.71 | 26.86 | 23.63 | 23.80 | 22.77 | 23.24 | 22.41 | 29.78 | 30.24 |
| ARP [35] | 51.64 | 36.29 | 34.11 | 31.56 | 26.56 | 23.38 | 23.74 | 22.98 | 23.42 | 22.86 | 29.65 | 32.08 |
| CEC [49] | 54.19 | 37.33 | 35.44 | 32.42 | 27.41 | 24.26 | 24.74 | 23.92 | 24.41 | 23.66 | 30.78 | 30.53 |
| FACT [50] | 53.14 | 36.96 | 35.29 | 32.54 | 27.55 | 24.42 | 24.74 | 23.71 | 24.35 | 23.39 | 30.61 | 29.74 |
| PAN [33] | 53.18 | 37.03 | 35.32 | 32.63 | 27.57 | 24.39 | 24.67 | 23.62 | 24.22 | 23.25 | 30.59 | 29.93 |
| AMFO [37] | 54.30 | 37.75 | 35.56 | 33.01 | 27.73 | 24.75 | 24.95 | 24.13 | 24.96 | 24.03 | 31.12 | 30.27 |
| AISP [39] | 54.95 | 40.77 | 37.91 | 34.38 | **29.08** | 25.89 | **26.86** | 26.45 | **26.94** | 26.03 | 32.93 | 28.92 |
| Ours | **55.50** | **40.85** | **39.97** | **34.60** | 27.04 | **28.95** | 24.96 | **27.23** | 26.03 | **27.55** | **33.27** | **27.95** |

### 3.4. Accuracy Comparison of Different Methods

The prior methods include the Finetune [47], iCaRL [48], DFSL [32], ARP [35], CEC [49], FACT [50], PAN [33], AMFO [37] and AISP [39]. The Finetune method adjusts model's parameters using samples of new classes. The iCaRL method retains data of old classes and distills knowledge. The DFSL method uses an attentive weight generator and classifier of cosine similarity. The ARP method refines prototypes using dynamic relation projection. The CEC method uses a graph network for prototype update. The FACT method reserves embedding space to hold new classes. The PAN method updates classifier by an adaption network. The AMFO method uses discriminative and generalizable classifier and embeddings. The AISP method uses adaptive embedding fusion to improve model's stability and plasticity.

Tables 4, 5 and 6 present the results achieved by various methods on the LS-100, NSynth-100 and FSC-89, respectively. Our method exceeds all prior methods in AA and PD. The advantage of our method is attributed to the design of MEE and RRC. The MEE improves the representational ability of the embeddings by fusing multi-level features, while the RRC reduces the correlations between various dimensions of each embedding and the model's generalizability can be improved.

### 3.5. Complexity Comparison of Different Methods

The computational complexity and memory overhead of each method are gauged by Multiply-ACcumulate operations (MACs) and Number of Parameters (NP), respectively. The MACs and NP have been widely used in previous works.

Table 7 lists the values of MACs and NP of various methods. MACs values of our method are 19.54 billion on the LS-100, 5.16 billion on the NSynth-100 and 3.10 billion on the FSC-89, while the NP values of our method are 87.82 million on the LS-100, 87.69 million on the NSynth-100 and 87.57 million on the FSC-89. These MACs and NP values are higher than the counterparts of the methods of Finetune, iCaRL, DFSL and FACT, but lower than the counterparts of the methods of ARP, CEC, PAN, AMFO and AISP. Compared to the methods of Finetune, iCaRL, DFSL and FACT, our method has a fusion module with high complexity in the MEE, which leads to higher MACs and NP values. Compared to the attention modules in the methods of ARP, CEC, PAN, AMFO and AISP, the fusion module used in our method has lower computational complexity and memory overhead. From the above results, it is known that the complexity of our method is in the middle position among all methods. In addition, the Finetune and iCaRL methods have the same lowest MACs and NP values. The reason is that their classifiers are a simple FC layer and do not expand with the increase of classes.

Table 7: *Values of MACs and NP of various methods*

| Methods | MACs (billion) | | | NP (million) | | |
|---|---|---|---|---|---|---|
| | LS-100 | NSynth-100 | FSC-89 | LS-100 | NSynth-100 | FSC-89 |
| Finetune | **19.45** | **5.12** | **3.07** | **86.97** | **86.84** | **86.82** |
| iCaRL | **19.45** | **5.12** | **3.07** | **86.97** | **86.84** | **86.82** |
| DFSL | 19.51 | 5.14 | 3.08 | 87.64 | 87.51 | 87.49 |
| ARP | 20.42 | 5.36 | 3.20 | 88.82 | 88.69 | 88.67 |
| CEC | 21.38 | 5.60 | 3.33 | 89.37 | 89.24 | 89.22 |
| FACT | 19.46 | 5.13 | 3.08 | 87.01 | 86.88 | 86.86 |
| PAN | 22.35 | 5.84 | 3.46 | 91.73 | 91.60 | 91.58 |
| AMFO | 21.56 | 5.78 | 3.40 | 92.53 | 92.40 | 92.38 |
| AISP | 38.91 | 10.24 | 6.14 | 174.63 | 174.37 | 174.33 |
| Ours | 19.54 | 5.16 | 3.10 | 87.82 | 87.69 | 87.57 |

## 4. Conclusions

In this work, we propose a FFCAC method using the MEE and RRC. Based on the experimental results and analyses, we can draw three conclusions. First, our method exceeds other methods in accuracy. Second, each main part of our method has contributions to the accuracy improvement. Third, the complexity of our method is in the middle of all methods. In the next work, we will improve the accuracy and reduce the complexity of our method by taking more effective measures, such as model distillation and compression.

## 5. Acknowledgements

This work was supported by the national natural science foundation of China (62371195, 62111530145, 61771200), the exchange project of the 10th Meeting of the China-Croatia Science and Technology Cooperation Committee (No. 10-34), and Guangdong S&T project (2023A0505050116), and Guangdong provincial key laboratory of human digital twin (2022B1212010004).